\documentclass[12pt]{article}
\begin{document}
\begin{center}
{\large\bf Charge Conjugation Invariance of the Vacuum and the
Cosmological Constant Problem} \vskip 0.3 true in {\large J. W.
Moffat} \vskip 0.3 true in {\it The Perimeter Institute for
Theoretical Physics, Waterloo, Ontario, N2J 2W9, Canada} \vskip
0.3 true in and \vskip 0.3 true in {\it Department of Physics,
University of Waterloo, Waterloo, Ontario N2Y 2L5, Canada}
\end{center}
%\date{\today}
\begin{abstract}%
We propose a method of field quantization which uses an indefinite
metric in a Hilbert space of state vectors. The action for gravity
and the standard model includes, as well as the positive energy
fermion and boson fields, negative energy fields. The Hamiltonian
for the action leads through charge conjugation invariance
symmetry of the vacuum to a cancellation of the zero-point vacuum
energy and a vanishing cosmological constant in the presence of a
gravitational field. To guarantee the stability of the vacuum, we
introduce a Dirac sea `hole' theory of quantization for gravity as
well as the standard model. The vacuum is defined to be fully
occupied by negative energy particles with a hole in the Dirac
sea, corresponding to an anti-particle. We postulate that the
negative energy bosons in the vacuum satisfy a para-statistics
that leads to a para-Pauli exclusion principle for the negative
energy bosons in the vacuum, while the positive energy bosons in
the Hilbert space obey the usual Bose-Einstein statistics. This
assures that the vacuum is stable for both fermions and bosons.
Restrictions on the para-operator Hamiltonian density lead to
selection rules that prohibit positive energy para-bosons from
being observable. The problem of deriving a positive energy
spectrum and a consistent unitary field theory from a
pseudo-Hermitian Hamiltonian is investigated.

\end{abstract}
\vskip 0.2 true in e-mail: john.moffat@utoronto.ca

%\pacs{ }

\section{Introduction}

The cosmological constant problem is considered to be one of the
major problems of modern physics~\cite{Weinberg}. The particle
physics origin of the problem arises because of the quartic
divergence of the zero-point vacuum energy in the presence of a
gravitational field. The constant zero-point energy cannot be
shifted to zero in the action due to the universal coupling of
gravity to energy including vacuum energy. It is hoped that a
natural symmetry exists that explains why the cosmological
constant is zero or small. However, the obvious candidates,
supersymmetry and conformal invariance, cannot supply this
solution, for they are badly broken in Nature. It is possible that
the standard second quantization method, that forms the basis for
modern quantum field theory, possesses a fundamental flaw in its
physical interpretation of the vacuum. Due to this possible faulty
understanding of the vacuum state in quantum field theory, we are
led to two important difficulties inherent in the theory: 1) a
generic infinity of the zero-point vacuum energy in the presence
of a gravitational field and, 2) the lack of renormalizability of
perturbative quantum gravity.

Linde~\cite{Linde,Linde2} suggested some time ago that that a
novel symmetry could be introduced between two different universes
with independent spacetimes. The symmetry transforms the action
$S$ into $-S$, so that the positive energy of one spacetime is
transformed into the negative energy of the other spacetime. The
action $S$ is symmetric under the transformation of the matter
field $\phi(x)\rightarrow {\overline\phi}(x)$ and the metric
$g_{\mu\nu}(x)\rightarrow {\overline g}_{\mu\nu}(x)$. A
consequence of this symmetry is the invariance under the change of
the values of the effective potentials $V(\phi)\rightarrow
V(\phi)+c$, $V(\overline\phi)\rightarrow V(\overline\phi)+c$,
where $c$ is a constant. However, in order to avoid negative
probabilities and the instability of the vacuum associated with
negative energy particles, Linde postulated that the positive and
negative energy fields do not interact. This postulate introduces
an extreme and unnatural fine-tuning of the model.

Recently, Kaplan and Sundrum~\cite{Sundrum} have introduced a
symmetry which transforms positive energy into negative energy
through a projection operator $P$. As in the case of Linde's two~-
universe proposal, the authors postulate that to avoid a breakdown
of the vacuum due to the negative energy particles, the two copies
of the standard model matter fields, corresponding to positive and
negative energy particles, interact only weakly through gravity.
To prevent excessively rapid decay of the vacuum, it is also
postulated that gravitational Lorentz invariance breaks down at
short distances.

In the following, we shall pursue a different approach to the
solution of the cosmological constant problem, and the possible
avoidance of the lack of renormalizability of perturbative quantum
gravity, by following a method of quantizing fields introduced by
Dirac in 1942~\cite{Dirac} and investigated in detail by
Pauli~\cite{Pauli} and Sudarshan and
collaborators~\cite{Sudarshan,Sudarshan2,Sudarshan3}. We
investigate a method of field quantization motivated by Bender and
collaborators~\cite{Bender,Bender2,Bender3,Bender4,Brandt,Bender5,Kleefeld,Znojil}
that can implement a positive energy spectrum for a
pseudo-Hermitian Hamiltonian, involving an indefinite metric in
Hilbert space, and a unitary $S$-matrix. A para-statistics is
invoked for negative energy bosons and their corresponding
``holes''in the vacuum, which together with the negative energy
Pauli exclusion principle for fermions and a para-Pauli exclusion
principle for negative energy bosons, leads to a stability of the
vacuum.

The charge conjugation invariance of the vacuum state leads to a
cancellation of the zero-point vacuum energy in the presence of a
gravitational field and to the vanishing of the cosmological
constant.

\section{Indefinite Metric in Hilbert Space}

Dirac~\cite{Dirac,Pauli} generalized field quantization by
introducing an indefinite metric in the Hilbert space of the state
vectors. The normalization of a state vector $\Psi$ is normally
defined by
\begin{equation}
\label{positivenorm} {\cal N}_+=\int dq\Psi^*\Psi,
\end{equation}
where $\Psi^*$ is the complex conjugate of $\Psi$. The scalar
product of two complex state vectors $\Phi$ and $\Psi$ is given by
\begin{equation}
{\cal B}_+=\int dq\Phi^*\Psi.
\end{equation}
Instead, we consider the more general bilinear form
\begin{equation}
{\cal B}=\int dq\Phi^*\eta\Psi,
\end{equation}
in which the operator $\eta$ is an Hermitian operator to guarantee
real normalization values.

The expectation value of an observable ${\cal O}$ described by a
linear operator is now defined by
\begin{equation}
\langle{\cal O}\rangle=\int dq\Psi^*\eta{\cal O}\Psi.
\end{equation}
The generalization of the standard Hermitian conjugate operator
${\cal O}={\cal O}^\dagger$ is given by the adjoint operator
\begin{equation}
\tilde{\cal O}={\eta}^{-1}{\cal
O}^\dagger\eta^\dagger={\eta}^{-1}{\cal O}^\dagger\eta.
\end{equation}
All physical observables have to be self-adjoint, $\tilde{\cal
O}={\cal O}$, to guarantee that their expectation values are real.
In particular, the Hamiltonian operator $H$ has to be
self-adjoint, $\tilde H=H$, which has the consequence that
\begin{equation}
\frac{d}{dt}\int dq\Psi^*\eta\Psi=i\Psi^*\eta (\tilde H-H)\Psi=0,
\end{equation}
guaranteeing the conservation of the normalization with time.

A linear transformation in the Hilbert space
\begin{equation}
\Psi'={\cal S}\Psi,
\end{equation}
requires that
\begin{equation}
\eta'={\cal S}^\dagger\eta {\cal S}.
\end{equation}
This assures that the normalization of the state vector is
invariant
\begin{equation}
\int dq\Psi^{*'}\eta'\Psi'=\int dq\Psi^*\eta\Psi.
\end{equation}
Then we have
\begin{equation}
{\cal O}'={\cal S}^{-1}{\cal O}{\cal S},\quad \tilde{\cal O}'
=\eta'^{-1}{\cal O}^{\dagger'}\eta'={\cal S}^{-1}\tilde{\cal
O}{\cal S},
\end{equation}
and the expectation value of the operator ${\cal O}$ is invariant
\begin{equation}
\int dq\Psi^{*'}\eta'{\cal O}'\Psi'=\int dq \Psi^*\eta{\cal
O}\Psi.
\end{equation}
We note that the usual Hermitian property of an operator is not
invariant with respect to the ${\cal S}$ transformation, whereas
the quality of being self-adjoint is invariant.

A transformation of the Hermitian matrix $\eta$ to normal,
diagonal form can have the values $1$ or $-1$. The positive
definite form (\ref{positivenorm}) yields a unit matrix and
positive probabilities. However, in general positive eigenvalues
can have negative probabilities, i.e. one can introduce negative
probabilities that certain positive eigenvalues of an observable
are realized. We shall discuss the possibility of obtaining a
physical interpretation of quantum field theory in which the
$S$-matrix is unitary and only a positive energy spectrum is
observed when the pseudo-Hermitian Hamiltonian is quantized.

\section{The Action and Field Quantization}

The action takes the form $(c=\hbar=1)$:
\begin{equation}
\label{gravaction} S=S_{\rm Grav}+S_M(\phi)+S_M(\chi),
\end{equation}
where
\begin{equation}
S_{\rm Grav}=\frac{1}{16\pi G}\int d^4x\sqrt{-g}[(R-2\Lambda_0)
-({\overline R}-2\overline\Lambda_0)].
\end{equation}
The $R$ denotes the normal Ricci scalar associated with positive
energy gravitons, while ${\overline R}$ is associated with
negative energy gravitons. Moreover, $\Lambda_0$ and
$\overline\Lambda_0$ denote the ``bare'' cosmological constants
corresponding to positive and negative energy gravitons. The
$\phi$ and $\chi$ fields denote positive and negative energy
matter fields, respectively.

We define an effective cosmological constant
\begin{equation}
\label{effectivelambda} \label{efflambda} \Lambda_{\rm
eff}=\Lambda_{0{\rm eff}}+\Lambda_{\rm vac},
\end{equation}
where $\Lambda_{0{\rm eff}}=\Lambda_0-\overline{\Lambda}_0$,
$\Lambda_{\rm vac}=8\pi G\rho_{\rm vac}$ and $\rho_{\rm vac}$
denotes the vacuum density.

We expand the metric tensor $g_{\mu\nu}$ about Minkowski flat
space
\begin{equation}
g_{\mu\nu}=\eta_{\mu\nu}+h_{\mu\nu}+O(h^2),
\end{equation}
where $\eta_{\mu\nu}={\rm diag}(1,-1,-1,-1)$. We begin with the
lowest weak field approximation for which $\sqrt{-g}=1$.

Let us consider as a first simple case a real scalar field
$\phi(x)$ in the absence of interactions. The action is
\begin{equation}
\label{phiaction} S_\phi=\frac{1}{2}\int
d^4x\biggl[\partial_\mu\phi\partial^\mu\phi-\mu^2\phi^2\biggr],
\end{equation}
and $\phi$ satisfies the wave equation
\begin{equation}
\label{waveeq} (\partial^\mu\partial_\mu+\mu^2)\phi=0.
\end{equation}
As is well known, this equation has both positive and negative
energy solutions, as is the case with the Dirac
equation~\cite{Bjorken}.

The Hamiltonian for positive energy scalar particles is given by
 \begin{equation}
 H=\frac{1}{2}\int
 d^3x\biggl[({\vec\nabla}\phi)^2+(\partial_0\phi)^2+\mu^2\phi^2\biggr].
 \end{equation}
 The usual decomposition of the field $\phi$ into Fourier
 components is given by
 \begin{equation}
 \phi(x)=V^{-1/2}\sum_k(2k_0)^{-1/2}\{\phi(k)\exp[i(\vec{k}\cdot{\vec{x}}-k_0x_0)]
 +\phi^*(k)\exp[i(-\vec{k}\cdot\vec{x}+k_0x_0)]\},
 \end{equation}
 where $k_0$ is positive, $k_0=+(\vec{k}^2+\mu^2)^{1/2}$. The
 standard quantization is based on the commutator
 \begin{equation}
 [\phi(k),\phi^*(k)]=1,
 \end{equation}
 and gives for the energy
 \begin{equation}
 E=\sum_kk_0\biggl[\frac{1}{2}+N_k\biggr].
 \end{equation}
 For the vacuum (ground state), $N_k=0$ and we obtain the zero-point
 vacuum energy
 \begin{equation}
 \label{zeropointenergy}
 E_0=\frac{1}{2}\sum_kk_0=\frac{1}{2(2\pi)^3}\int d^3k\sqrt{\vec{k}^2+\mu^2}.
 \end{equation}

 The zero-point vacuum energy $E_0$ diverges quartically and is
 the root of the cosmological constant problem in the presence of
 a higher-order gravitational field, since graviton loops can couple
 to the vacuum energy ``bubble'' graphs which cannot be
 time-ordered away, i.e. we cannot simply shift the constant
 vacuum energy, $E_0$, such that only $E'=E-E_0$ is observed.

 We shall follow Dirac and Pauli~\cite{Dirac,Pauli} and decompose the real scalar field $\phi$
 according to
 \begin{equation}
 \phi(x)=\frac{1}{\sqrt{2}}[A(x)+\tilde A(x)].
 \end{equation}
 The quantization of $A(x)$ with $-k_0x_0$ in the phase factor
 occurs in the usual way, corresponding to positive energy
 particles, while the other part with $+k_0x_0$ in the phase
 factor is quantized such that it leads to negative energy
 particles. We set
 \begin{equation}
 \label{Aequation}
 A(x)=V^{-1/2}\sum_k(2k_0)^{-1/2}\{A_+(k)\exp[i(\vec{k}\cdot\vec{x}-k_0x_0)]
 $$ $$
 +A_-(k)\exp[i(-\vec{k}\cdot\vec{x}+k_0x_0)]\},
 \end{equation}
 \begin{equation}
 \label{tildeAequation}
\tilde A(x) =V^{-1/2}\sum_k(2k_0)^{-1/2}\{\tilde
A_+(k)\exp[i(-\vec{k}\cdot\vec{x}+k_0x_0)]
$$ $$
 +\tilde A_-(k)\exp[i(\vec{k}\cdot\vec{x}-k_0x_0)]\}.
 \end{equation}

 We now define the Hamiltonian to be
 \begin{equation}
 \label{Ahamiltonian}
 H=\int d^3x[\vec\nabla \tilde{A}\vec\nabla A+\partial_0 \tilde A\partial_0
 A+\mu^2{\tilde A}A].
 \end{equation}
 We assume that $\tilde A_+(k)$, $A_+(k)$ commute with $\tilde A_-(k)$,
 $A_-(k)$ and set
 \begin{equation}
 [A_+(k),\tilde A_+(k)]=1,\quad [A_-(k),\tilde A_-(k)]=-1,
 \end{equation}
leaving the usual commutation relation for ${\tilde A}(x)$, $A(x)$
unchanged. We now obtain
\begin{equation}
N_+(k)=\tilde A_+(k)A_+(k),\quad N_-(k)=-{\tilde A}_-(k)A_-(k).
\end{equation}
This leads to the energy
\begin{equation}
E=\sum_kk_0\biggl[\biggl(\frac{1}{2}+N_+(k)\biggr)-\biggl(\frac{1}{2}
+N_-(k)\biggr)\biggr]=\sum_kk_0[N_+(k)-N_-(k)].
\end{equation}

We see that {\it the zero-point vacuum energy has cancelled} in
the expression for the total energy $E$. The underlying symmetry
that has cancelled the divergent $E_0$ is charge conjugation
invariance of the vacuum state $\vert 0\rangle$. Thus, we have
uncovered a symmetry in particle physics that cancels the
divergent zero-point vacuum energy. This will still hold in the
presence of interactions, in particular, it will hold in the
presence of the gravitational coupling of gravitons to the matter
fields. Moreover, we can show that for the quantized spin 2
graviton field, the composition of both positive and negative
energy gravitons leads to the cancellation of the zero-point
graviton vacuum energy and its contribution to $\Lambda_{\rm
vac}$.

An alternative quantization procedure consists of defining besides
the field $\phi(x)$ another scalar field $\chi(x)$, the adjoint of
which is $\tilde\chi(x)=-\chi(x)$. Then we have
\begin{equation}
\chi(x)=\frac{1}{\sqrt{2}}[A(x)-\tilde A(x)],
\end{equation}
with the Fourier decomposition
\begin{equation}
 \chi(x)=V^{-1/2}\sum_k(2k_0)^{-1/2}\{\tilde\chi(k)\exp[i(\vec{k}\cdot\vec{x}-k_0x_0)]
 -\chi(k)\exp[i(-\vec{k}\cdot\vec{x}+k_0x_0)]\}.
 \end{equation}
 The $\chi$ field is quantized according to
 \begin{equation}
 [\chi(k),\tilde\chi(k)]=-1,
 \end{equation}
 which gives
 \begin{equation}
 \tilde\chi(k)\chi(k)=-N_{\chi}(k).
 \end{equation}
 The Hamiltonian is now given by
 \begin{equation}
 H=\frac{1}{2}\int
 d^3x\biggl[(\vec\nabla\phi)^2+(\partial_0\phi)^2+\mu^2\phi^2
 -(\vec\nabla\chi)^2-(\partial_0\chi)^2-\mu^2\chi^2\biggr].
 \end{equation}
 This leads to the energy
 \begin{equation}
 E=\sum_kk_0[N_\phi-N_\chi].
 \end{equation}
 As before, the charge conjugation invariance of the vacuum state
 leads to the cancellation of the zero-point vacuum energy, $E_0$.
 We have
 \begin{equation}
 \phi(k)=\frac{1}{\sqrt{2}}[A_+(k)+\tilde A_-(k)],\quad
 \tilde\phi(k)=\frac{1}{\sqrt{2}}[\tilde A_+(k)+A_-(k)],
 \end{equation}
 \begin{equation}
 \chi(k)=\frac{1}{\sqrt{2}}[\tilde A_+(k)-A_-(k)],\quad
 \tilde\chi(k)=\frac{1}{\sqrt{2}}[A_+(k)-\tilde {A}_-(k)].
 \end{equation}

The normalization of the state vector $\Psi$ is given
by~\cite{Pauli}:
\begin{equation}
{\cal N}=\sum_{N_+(k),N_-(k)}(-1)^{\sum
N_-(k)}\Psi^*(...N_+(k)...,...N_-(k)...)
$$ $$
\times\Psi(...N_+(k)..., ...N_-(k)...)={\rm const.}
\end{equation}
This demonstrates that ``negative probability'' states will exist
with an odd number of particles in states with negative energy. We
shall see in the following section, how we can quantize the field
in the presence of interactions and avoid a catastrophic
instability due to these negative probabilities and negative
energy particles. In Section 5, we shall investigate how we can
formulate the quantum field theory, so that we obtain a real and
positive energy spectrum and a unitary $S$-matrix.

It can be shown that an equivalent quantization procedure for
complex charged scalar fields, neutral vector gauge fields
$A_\mu$, spin-2 graviton fields, and Dirac spinor fields can be
derived that leads to the cancellation of the zero-point vacuum
energy and the vanishing of $\Lambda_{\rm vac}$ in the presence of
a gravitational field. This is again due to the charge conjugation
invariance of the vacuum state $\vert 0\rangle$ for these fields.

\section{Charge Conjugation and Para-Statistics for
Negative Energy Bosons in the Vacuum}

A spinor field $\psi$ satisfies in the presence of an
electromagnetic interaction the Dirac equation
\begin{equation}
[\gamma^\mu (p_\mu-eA_\mu)-m]\psi=0,
\end{equation}
and its charge conjugate equation
\begin{equation}
[\gamma^\mu(p_\mu+eA_{c\mu})-m]\psi_c=0.
\end{equation}
These equations yield both positive and negative energy solutions
of the Dirac equation. The two spinor fields $\psi$ and $\psi_c$
and the two photon fields $A_\mu$ and $A_{c\mu}$ are associated
with positive and negative energy fermions and neutral gauge
fields, respectively. For the charge conjugation transformation we
have
\begin{equation}
\psi_c=C\gamma^0\psi^*=C{\overline\psi}^T,
\end{equation}
where $C$ is the charge conjugation matrix which satisfies
\begin{equation}
C^{-1}\gamma^\mu C=-\gamma^{\mu T},
\end{equation}
and $\gamma^{\mu T}$ denotes the transpose of the Dirac $\gamma$
matrix. A similar transformation exists for the gauge field
$A_\mu$ which transforms it into the anti-particle gauge field
(the photon is its own anti-particle).

Consider the periodic solutions of the positive energy Dirac
spinor field $\psi^+_\sigma(x)$:
\begin{equation}
\psi^+_\sigma(\vec{x})=\sum_m
a_m\exp(-itE_m)u^+_{m\sigma}(\vec{x}),\quad
\tilde\psi^+_\sigma(\vec{x})=\sum_m \tilde a_m\exp(itE_m)\tilde
u^+_{m\sigma}(\vec{x}),
\end{equation}
where
\begin{equation}
\int d^3x\tilde u_mu_{m'}=\delta_{mm'}.
\end{equation}
The canonical momentum $\pi^+_\sigma(\vec{x})$ is
\begin{equation}
\pi^+_\sigma(\vec{x})=i\sum_m \tilde a_m\exp(itE_m)\tilde
u^+_{m\sigma}(\vec{x}).
\end{equation}
Then, the anti-commutator quantization is
\begin{equation}
\{\psi_\sigma^+(\vec{x},0),\pi^+_{\sigma'}(\vec{x'},0)\}=i\delta_{\sigma\sigma'}
\delta(\vec{x}-\vec{x'}).
\end{equation}

We now introduce a negative energy spinor field
$\psi^-_\sigma(x)$:
\begin{equation}
\psi^-_\sigma(\vec{x})=\sum_m
b_m\exp(itE_m)u^-_{m\sigma}(\vec{x}),\quad
\tilde\psi^-_\sigma(\vec{x})=\sum_m \tilde b_m\exp(-itE_m)\tilde
u^-_{m\sigma}(\vec{x}).
\end{equation}
The canonical momentum is given by
\begin{equation}
\pi^-_\sigma(\vec{x})=i\sum_m \tilde b_m\exp(-itE_m)\tilde
u^-_{m\sigma}(\vec{x}),
\end{equation}
and the anti-commutation rule is
\begin{equation}
\{\psi^-_{\sigma}(\vec{x},0),\pi^-_{\sigma'}(\vec{x'},0)\}=-i\delta_{\sigma\sigma'}
\delta(\vec{x}-\vec{x'}).
\end{equation}
The total energy is given by
\begin{equation}
E=\sum_mE_m[N^+_m-N_m^-].
\end{equation}
The zero-point vacuum energy has cancelled for the fermions.

The very existence of negative energy solutions for electrons
motivated Dirac to introduce his ``hole'' theory~\cite{Dirac2}.
This theory prevents a catastrophic instability of the vacuum,
which prevents atomic electrons from making radiative transitions
into negative energy states. The rate at which electrons make a
transition into the negative energy interval $-m$ to $-2m$ is
\begin{equation}
{\cal R}\sim \frac{2\alpha^6m}{\pi}\sim 10^8\,{\rm sec}^{-1}.
\end{equation}
The rate blows up if all the negative energy levels are taken into
account. The stability of the hydrogen ground state is guaranteed
by using the Pauli exclusion principle. All the negative energy
fermion levels of the vacuum are filled up and all positive energy
levels are empty. The stability of the hydrogen-atom ground state
is assured, since no more electrons can be accommodated in the
negative energy sea by the Pauli exclusion principle. A ``hole''
in the negative energy sea of fermions is now an anti-particle
fermion (for the electron it is a positron). It is possible for a
negative energy fermion to absorb radiation and be excited into a
positive energy state, when we observe a fermion of charge $-\vert
e\vert$ and energy $+E$ and in addition a hole in the negative
energy sea leading to pair production. If we denote the energy
levels of positive and negative energies by $E >0$ and $E <0$,
respectively, then according to Fermi-Dirac statistical weights
the occupation numbers are restricted to the values $0$ or $1$.
Then, the vacuum is described by all negative energy levels being
occupied and all positive energy levels being empty: $N_{fm}=1$
for $E_{fm} < 0$ and $N_{fm}=0$ for $E_{fm} > 0$.

In standard second quantized field theory and the interpretation
of current particle physics, we obtain a consistent particle
quantum field theory by interpreting creation and annihilation of
a negative energy particle as the annihilation and creation of a
positive energy particle. The negative energy particle vacuum is
empty as in the vacuum of a positive energy solution.\footnote{A
negative energy boson vacuum Dirac sea has been investigated by
Habara, Nielsen and Ninomiya, and compared with a supersymmetric
point of view~\cite{Nielsen}.}

We must now also assure the stability of the vacuum against the
cascade of positive energy bosons into negative energy states and
then into the ground state. We do this by postulating that
negative energy bosons in the vacuum satisfy para-statistics. All
the negative energy para-boson levels in the vacuum are filled and
all the positive energy para-boson levels are empty. The negative
energy bosons including the photon and the graviton obey a
para-Pauli exclusion principle. We now have for the bosons the
energy
\begin{equation}
E=\sum_kk_0\biggl[\biggl(\frac{1}{2}+N^+_{bk}\biggr)
-\biggl(\frac{1}{2}+N^-_{bk}\biggr)\biggr]
=\sum_kk_0[N^+_{bk}-N^-_{bk}].
\end{equation}
As before, the zero-point vacuum energy has cancelled. However,
due to the boson negative energy, para-statistics the boson part
of the vacuum is stable.

Let us assume that a field theory satisfies the properties:
\begin{enumerate}

\item It is invariant under proper Lorentz transformations,

\item Its Hamiltonian is Hermitian, $H^\dagger=H$,

\item The usual connection between spin and statistics is made,

\item The field operators are local.

\end{enumerate}

Then it can be shown that such a field theory is invariant under
the product operation $TCP$~\cite{Luders,Pauli2,Streater}. We
observe that in our field theory formulation our Hamiltonian is
not Hermitian but satisfies the adjoint symmetry, $\tilde H=H$. We
may expect that TCP invariance and Lorentz invariance will be
broken at some high scale of energy. However, this is an issue
that requires further investigation.

The interpretation of the filled negative energy boson sea in the
vacuum is based on the para-statistics introduced by
Green~\cite{Green} and studied further by Greenberg and
Messiah~\cite{Greenberg}, Dr\"uhl, Haag and Roberts~\cite{Haag}
and Streater and Wightman~\cite{Streater}. A para-boson field is
defined to be of order $p$:
\begin{equation}
\phi(x)=\sum_{i=1}^p\phi^{(i)}(x),
\end{equation}
where if $x$ and $y$ are spacelike separated $(x-y)^2 < 0$, then
we have
\begin{equation}
[\phi^{(i)}(x),\phi^{(i)}(y)]=0,\quad
\{\phi^{(i)}(x),\phi^{(j)}(y)\}=0\,\,{\rm if}\,\, i\not= j.
\end{equation}
The para-statistics field theory for bosons does not agree with
the L\"uders-Pauli statistics theorem. When a positive energy
boson satisfying normal Bose-Einstein statistics tries to transit
into a negative energy state, a para-Pauli exclusion principle
prohibits this from happening. It was shown by Greenberg and
Messiah~\cite{Greenberg} that restrictions on the form of the
interaction Hamiltonian density, $H_I$, are derived by requiring
that $H_I$ be a para-operator. From this restriction on $H_I$,
selection rules are derived for the $S$-matrix which prohibit all
reactions in which the total number of para particles of order $p
> 1$ in the initial and final states is 1. This selection
rule together with experimental input, leads to the conclusion
that para-bosons with positive energy will not be observed. This
picture of filled negative energy para-bosons in the vacuum and
the corresponding positive energy para-anti-bosons will assure
that the boson part of the vacuum is stable against catastrophic
collapse.

The negative energy gravitons will also fill a sea of negative
energy levels in the vacuum corresponding to gravitons obeying
para-statistics. Selection rules for the graviton-graviton and
graviton-matter interactions will prohibit positive energy
para-gravitons from being observable.

\section{Pseudo-Hermitian Hamiltonian and Unitarity}

In standard quantum field theory the Hamiltonian is Hermitian,
$H^\dagger=H$, and we are assured that the energy spectrum is real
and that the time evolution of the operator $U=\exp(itH)$ is
unitary and probabilities are positive and preserved for particle
transitions. However, in recent years there has been a growth of
activity in studying quantum theories with pseudo-Hermitian
Hamiltonians, which satisfy the generalized property of
adjointness, $\tilde H=\eta^{-1}H^\dagger\eta$, associated with an
indefinite metric in Hilbert
space~\cite{Bender,Bender2,Bender3,Bender4,Brandt,Bender5,Kleefeld,Znojil}.

Spectral positivity and unitarity can in special circumstances
follow from a symmetry property of the Hamiltonian in terms of the
symmetry under the operation of ${\cal P}{\cal T}$, where ${\cal
P}$ is a linear operator represented by parity reflection, while
${\cal T}$ is an anti-linear operator represented by time
reversal. If a Hamiltonian has an unbroken ${\cal P}{\cal T}$
symmetry, then the energy levels can in special cases be real and
the theory can be unitary and free of ``ghosts''. The operation of
${\cal P}$ leads to $\vec{x}\rightarrow -\vec{x}$, while the
operation of ${\cal T}$ leads to $i\rightarrow -i$ (or
$x^0\rightarrow -x^0$). It follows that under the operation of
${\cal P}{\cal T}$ the Hamiltonian $H$ in (\ref{Ahamiltonian}) is
invariant under the ${\cal P}{\cal T}$ transformation, provided
$\tilde H=H$ and $A_+\tilde A_-=A_-\tilde A_+$, which is necessary
but not sufficient to assure the reality of the energy
eigenvalues.

The proof of unitarity follows from the construction of a linear
operator ${\cal C}$. This operator is used to define the inner
product of state vectors in Hilbert space:
\begin{equation}
\label{innerprod} \langle\Psi\vert\Phi\rangle=\Psi^{{\cal C}{\cal
P}{\cal T}}\cdot\Phi.
\end{equation}
Under general conditions, it can be shown that a necessary and
sufficient condition for the existence of the inner product
(\ref{innerprod}) is the reality of the energy spectrum of
$H$~\cite{Mostafazadeh,Swanson}. With respect to this inner
product, the time evolution of the quantum theory is unitary. In
quantum mechanics and in quantum field theory, the operator ${\cal
C}$ has the general form
\begin{equation}
{\cal C}=\exp(Q){\cal P},
\end{equation}
where $Q$ is a function of the dynamical field theory variables.
The form of ${\cal C}$ must be determined by solving for the
function $Q$ in terms of chosen field variables and field
equations. The form of ${\cal C}$ has been calculated for several
simple field theories, e.g. $\phi^3$ theory and also in massless
quantum electrodynamics with a pseudo-Hermitian Hamiltonian. The
solution for ${\cal C}$ satisfies
\begin{equation}
{\cal C}^2=1,\quad [{\cal C},{\cal P}{\cal T}]=0,\quad [{\cal
C},H]=0.
\end{equation}
We shall not attempt to determine a specific generalized charge
conjugation operator ${\cal C}$ in the present work.

It has also been shown that a special form of ${\cal P}$ leads to
a Lorentz invariant scalar expression for the operator ${\cal
C}$~\cite{Brandt}.

\section{The Resolution of the Cosmological Constant Problem}

We shall postulate that
\begin{equation}
\label{barelambda} \Lambda_{0{\rm
eff}}=\Lambda_0-{\overline\Lambda}_0=0.
\end{equation}
The vanishing of the zero-point vacuum energy in our quantum field
theory, including higher-order graviton tree-graph couplings and
loops, then assures that (\ref{barelambda}) is protected against
all higher order radiative vacuum corrections.

If we assume that a spontaneous symmetry breaking of the charge
conjugation invariance (or ${\cal C}$ invariance of the vacuum)
occurs, then this will create a small `` observed'', effective
cosmological constant $\Lambda_{\rm eff}/8\pi G\sim (2\times
10^{-3}\,eV)^4$, needed to provide a cosmological constant
explanation of the accelerating expansion of the
universe~\cite{Perlmutter,Riess,Spergel}. However, it is possible
that the accelerating expansion of the universe can be explained
by a late-time inhomogeneous cosmological model~\cite{Moffat}, in
which the cosmological constant $\Lambda_{\rm eff}=0$ and there is
no need for a negative pressure ``dark energy''.

A Casimir vacuum energy has been experimentally observed. In our
quantum field theory, the vanishing of the zero-point vacuum
energy is only valid in the absence of material boundary
conditions as are necessary for the Casimir effect~\cite{Jaffe}.
When material boundary conditions such as the parallel metal
plates required to perform the Casimir experiments are imposed,
then we can no longer demand that the generalized charge
conjugation ${\cal C}$ invariance of the vacuum state is
preserved; the breaking of {\cal C} invariance of the vacuum will
produce a non-vanishing zero-point energy effect.

Jaffe~\cite{Jaffe} points out that the Casimir effect gives no
more or less evidence for the ``reality'' of the vacuum
fluctuation energy of quantum fields than any other one-loop
effect in quantum electrodynamics, e.g. the vacuum polarization
effect associated with charges and currents in atomic physics.
Like all other observable effects in quantum electrodynamics, the
Casimir effect vanishes as the fine structure constant $\alpha$
goes to zero.

\section{Conclusions}

We have formulated a quantum field theory based on an indefinite
metric in Hilbert space with a generalization of the Hermitian
operator $H=H^\dagger$ to an adjoint operator $\tilde
H=\eta^{-1}H^\dagger\eta$ and we have ${\tilde H}=H$. The
quantization of fields in the presence of gravity is performed
with a positive and negative energy particle interpretation, which
leads to the cancellation of the zero-point vacuum energy due to
the generalized charge conjugation ${\cal C}$ invariance of the
vacuum:
\begin{equation}
\label{vacH} \langle 0\vert H\vert 0\rangle=0.
\end{equation}
We postulate that the effective, classical  ``bare'' cosmological
constant $\Lambda_{0{\rm eff}}=0$. The condition (\ref{vacH})
leads to a protection of the vanishing of the effective
cosmological constant, $\Lambda_{\rm eff}$, from all higher order
gravitational and external field quantum corrections.

It is assumed that all negative energy levels of fermions are
filled in the vacuum, according to Dirac's negative energy sea
theory, so that a catastrophic instability of the fermion part of
the vacuum is prevented due to the Pauli exclusion principle. To
prevent a similar catastrophic instability of the boson part of
the vacuum, we postulate that the negative energy bosons in the
vacuum satisfy para-statistics, whereby all the negative
para-boson energy levels in the vacuum are filled and all positive
energy para-boson levels are empty. A para-Pauli exclusion
principle prevents an instability of the boson part of the vacuum.
The positive energy normal bosons satisfy the usual Bose-Einstein
commutator quantization rule. It is possible to introduce the idea
of supersymmetry into our proposed Dirac hole theory of bosons and
fermions, for supersymmetric bosons and fermions are treated on
the same footing. In the present work we have kept to a
non-supersymmetric interpretation.

The indefinite Hilbert space state vector metric can generatef
negative probabilities for the transitions of particles and
violate the unitarity of the $S$-matrix. To guarantee positive
probabilities and the unitarity of transition and scattering
amplitudes, we incorporate the ${\cal P}{\cal T}$ operation on
field operators and the action. The generalized charge conjugation
operator ${\cal C}$, introduced by Bender and
collaborators~\cite{Bender,Bender2,Bender3,Bender4,Bender5} is
invoked to guarantee that the energy spectrum for gravity and the
standard model particle theory is positive, and assure that
probabilities are positive and conserved and that the $S$-matrix
is unitary.

The possibility that the new quantum field theory can lead to a
perturbatively renormalizable quantum gravity theory will be
investigated in a future publication.

\vskip 0.2 true in{\bf Acknowledgment} \vskip 0.2 true in

This research was supported by the Natural Sciences and
Engineering Research Council of Canada. I thank Christopher
Beetle, Joel Brownstein, Martin Green and Stefan Hofmann for
helpful and stimulating discussions.

\end{document}